\documentclass[twocolumn,superscriptaddress,showpacs,amssymb,preprintnumbers,prl,floatfix]{revtex4-1}
\usepackage{epsfig,dcolumn}
\usepackage{hyperref}
\usepackage{multirow}
\usepackage{bm,color}
\usepackage{tikz}
\usepackage{makecell}
\usepackage{graphicx,bm}
\newcommand{\be}{\begin{equation}}
	\newcommand{\ee}{\end{equation}}
\newcommand{\ba}{\begin{eqnarray*}}
	\newcommand{\ea}{\end{eqnarray*}}

\usepackage{ulem}

\usepackage{soul}

\begin{document}
	\title{$N=16$ magicity revealed at the proton drip-line through the study of $^{35}$Ca}

	\author{L.~Lalanne}
	\email[]{louis-alexandre.lalanne@cern.ch}
	\affiliation{Universit\'e Paris-Saclay, CNRS/IN2P3, IJCLab, 91405 Orsay, France}
	\affiliation{Grand Accélérateur National d'Ions Lourds (GANIL), CEA/DRF-CNRS/IN2P3, 
		Bd. Henri Becquerel, 14076 Caen, France}  
	\author{O.~Sorlin}
	\email[]{olivier.sorlin@ganil.fr}
	\affiliation{Grand Accélérateur National d'Ions Lourds (GANIL), CEA/DRF-CNRS/IN2P3, 
		Bd. Henri Becquerel, 14076 Caen, France}
	\author{A.~Poves}
	\affiliation{Departamento de F\'isica Te\'orica and IFT-UAM/CSIC, Universidad Aut\'onoma de Madrid,  E-2804 Madrid, Spain}
	\author{M.~Assi\'e}
	\affiliation{Universit\'e Paris-Saclay, CNRS/IN2P3, IJCLab, 91405 Orsay, France}
	\author{F.~Hammache}
	\affiliation{Universit\'e Paris-Saclay, CNRS/IN2P3, IJCLab, 91405 Orsay, France}
	\author{S.~Koyama}
	\affiliation{Department of Physics, Unviversity of Tokyo}
	\affiliation{Grand Accélérateur National d'Ions Lourds (GANIL), CEA/DRF-CNRS/IN2P3, 
		Bd. Henri Becquerel, 14076 Caen, France}
	\author{D.~Suzuki}
	\affiliation{RIKEN Nishina Center, 2-1, Hirosawa, Wako, Saitama 351-0198, Japan}
	\author{F.~Flavigny}
	\affiliation{Normandie Univ, ENSICAEN, UNICAEN, CNRS/IN2P3, LPC Caen, 14000 Caen, France}
	\author{V.~Girard-Alcindor}
	\affiliation{Grand Accélérateur National d'Ions Lourds (GANIL), CEA/DRF-CNRS/IN2P3, 
		Bd. Henri Becquerel, 14076 Caen, France}
	\author{A.~Lemasson}
	\affiliation{Grand Accélérateur National d'Ions Lourds (GANIL), CEA/DRF-CNRS/IN2P3, 
		Bd. Henri Becquerel, 14076 Caen, France}
	\author{A.~Matta}
	\affiliation{Normandie Univ, ENSICAEN, UNICAEN, CNRS/IN2P3, LPC Caen, 14000 Caen, France}
	\author{T.~Roger}
	\affiliation{Grand Accélérateur National d'Ions Lourds (GANIL), CEA/DRF-CNRS/IN2P3, 
		Bd. Henri Becquerel, 14076 Caen, France}
	\author{D.~Beaumel}
	\affiliation{Universit\'e Paris-Saclay, CNRS/IN2P3, IJCLab, 91405 Orsay, France}
	\author{Y~Blumenfeld}
	\affiliation{Universit\'e Paris-Saclay, CNRS/IN2P3, IJCLab, 91405 Orsay, France}
	\author{B.~A.~Brown}
	\affiliation{Department of Physics and Astronomy, National Superconducting Cyclotron Laboratory,
		Michigan State University, East Lansing, Michigan}
	\author{F.~De Oliveira Santos}
	\affiliation{Grand Accélérateur National d'Ions Lourds (GANIL), CEA/DRF-CNRS/IN2P3,
		Bd. Henri Becquerel, 14076 Caen, France}
	\author{F.~Delaunay}
	\affiliation{Normandie Univ, ENSICAEN, UNICAEN, CNRS/IN2P3, LPC Caen, 14000 Caen, France}
	\author{N.~de~S\'{e}r\'{e}ville}
	\affiliation{Universit\'e Paris-Saclay, CNRS/IN2P3, IJCLab, 91405 Orsay, France}
	\author{S.~Franchoo}
	\affiliation{Universit\'e Paris-Saclay, CNRS/IN2P3, IJCLab, 91405 Orsay, France}
	\author{J.~Gibelin}
	\affiliation{Normandie Univ, ENSICAEN, UNICAEN, CNRS/IN2P3, LPC Caen, 14000 Caen, France}
	\author{J.~Guillot}
	\affiliation{Universit\'e Paris-Saclay, CNRS/IN2P3, IJCLab, 91405 Orsay, France}
	\author{O.~Kamalou}
	\affiliation{Grand Accélérateur National d'Ions Lourds (GANIL), CEA/DRF-CNRS/IN2P3, 
		Bd. Henri Becquerel, 14076 Caen, France}
	\author{N.~Kitamura}
	\affiliation{Center for Nuclear Study, University of Tokyo}
	
	\author{V.~Lapoux}
	\affiliation{CEA, Centre de Saclay, IRFU, Service de Physique Nucléaire, 91191 Gif-sur-Yvette, France}
	
	\author{B.~Mauss}
	\affiliation{RIKEN Nishina Center, 2-1, Hirosawa, Wako, Saitama 351-0198, Japan}
	\affiliation{Grand Accélérateur National d'Ions Lourds (GANIL), CEA/DRF-CNRS/IN2P3, 
		Bd. Henri Becquerel, 14076 Caen, France}
	\author{P.~Morfouace}
	\affiliation{Grand Accélérateur National d'Ions Lourds (GANIL), CEA/DRF-CNRS/IN2P3, 
		Bd. Henri Becquerel, 14076 Caen, France}
	
	\author{J.~Pancin}
	\affiliation{Grand Accélérateur National d'Ions Lourds (GANIL), CEA/DRF-CNRS/IN2P3, 
		Bd. Henri Becquerel, 14076 Caen, France}
	
	\author{T.~Y.~Saito}
	\affiliation{Department of Physics, University of Tokyo}
	
	\author{C.~Stodel}
	\affiliation{Grand Accélérateur National d'Ions Lourds (GANIL), CEA/DRF-CNRS/IN2P3, 
		Bd. Henri Becquerel, 14076 Caen, France}
	\author{J-C.~Thomas}
	\affiliation{Grand Accélérateur National d'Ions Lourds (GANIL), CEA/DRF-CNRS/IN2P3, 
		Bd. Henri Becquerel, 14076 Caen, France}

	\begin{abstract}
		
		The last proton bound calcium isotope $^{35}$Ca has been studied for the first time, using the $^{37}$Ca$(p,t)^{35}$Ca two neutron transfer reaction. The radioactive $^{37}$Ca nuclei, produced by the LISE spectrometer at GANIL, interacted with the protons of the liquid hydrogen target CRYPTA, to produce tritons $t$ that were detected in the MUST2 detector array, in coincidence with the heavy residues Ca or Ar. The atomic mass of $^{35}$Ca and the energy of its first 3/2$^+$ state are reported. A large $N=16$ gap of 4.61(11)~MeV is deduced from the mass measurement, which together with other measured properties, makes $^{36}$Ca a doubly-magic nucleus. The $N=16$ shell gaps in $^{36}$Ca and $^{24}$O are of similar amplitude, at both edges of the valley of stability. This feature is discussed in terms of nuclear forces involved, within state-of-the-art shell model calculations. Even though the global agreement with data is quite convincing, the calculations underestimate the size of the $N=16$ gap in $^{36}$Ca  by 840(110) keV. 
		
	
\end{abstract}


\date{\today}
\maketitle

\noindent {\sl Introduction.} 
Magic nuclei, corresponding to special numbers of neutrons and/or protons for which shell gaps are large, feature an enhanced stability as compared to others. Some of the ``classical" magic numbers, well identified in stable nuclei (2, 8, 20, 28, 50, 82 and 126), are found to collapse in exotic regions of the chart of nuclides under the combined actions of nuclear forces and correlations (see, e.g. \cite{Sorl08, Gade15, Otsu20, Nowa21}).  In fact, in the last decades, many experimental and theoretical efforts have proven the disappearance of magic numbers in neutron rich nuclei such as $N=8$ \cite{Iwas00, Pain06, Imai09, Krie12,  Meha12, Mors18, Chen18}, $N=20$ \cite{Guil84, Moto95, Iwas01, Yana03, Terr08, Craw16, Door16, Lica19} or $N=28$ \cite{Bast07, Forc10, Take12, Sorl13, Rile19, Long21} and also the appearance of new shell closures such as $N=16$ \cite{Ozaw00, Kanu09, Hoff09, Tsho12}, $N=32$ \cite{Huck85, Wien13,Rose15} or $N=34$ \cite{Step13, Mich18, Iimu23}. 

Considering the recently discovered $N=32$ and $34$ neutron (sub-)shell closures \cite{Huck85, Wien13,Rose15,Step13, Mich18, Iimu23} together with the well established ones at $N=20$ and $N=28$, the Ca isotopes exhibit thus far the largest number of identified magic nuclei within a single isotopic chain. On the neutron-deficient side of the Ca chain, the high excitation energy of the first 2$^+$ and 1$^+$ states in $^{36}$Ca, as well as their large neutron-removal spectroscopic factors \cite{Lal22}, point to a significant sub-shell closure at $N=16$, the size of which remains to be determined. 

The magic number $N=16$ has been identified only around $^{24}$O until now, in replacement of $N=20$ \cite{Otsu01}, which  vanishes around $^{28}$O \cite{Reve20, Bagc20}. This $N=16$ magicity was inferred from the combined information of the drop in interaction cross sections \cite{Ozaw00}, the high excitation energy of the first 2$^+$ state in $^{24}$O \cite{Hoff09} and its small quadrupole deformation  \cite{Tsho12}. 

In this letter, we report on the first measurement of the atomic mass and first excited state of the last proton bound Ca isotope $^{35}$Ca, produced in the $^{37}$Ca$(p,t)^{35}$Ca reaction, evidencing the magicity of $N=16$ close to  the proton drip-line. 

\begin{figure}[!h]
\begin{center}
	\centering
	\includegraphics[width=0.99\columnwidth]{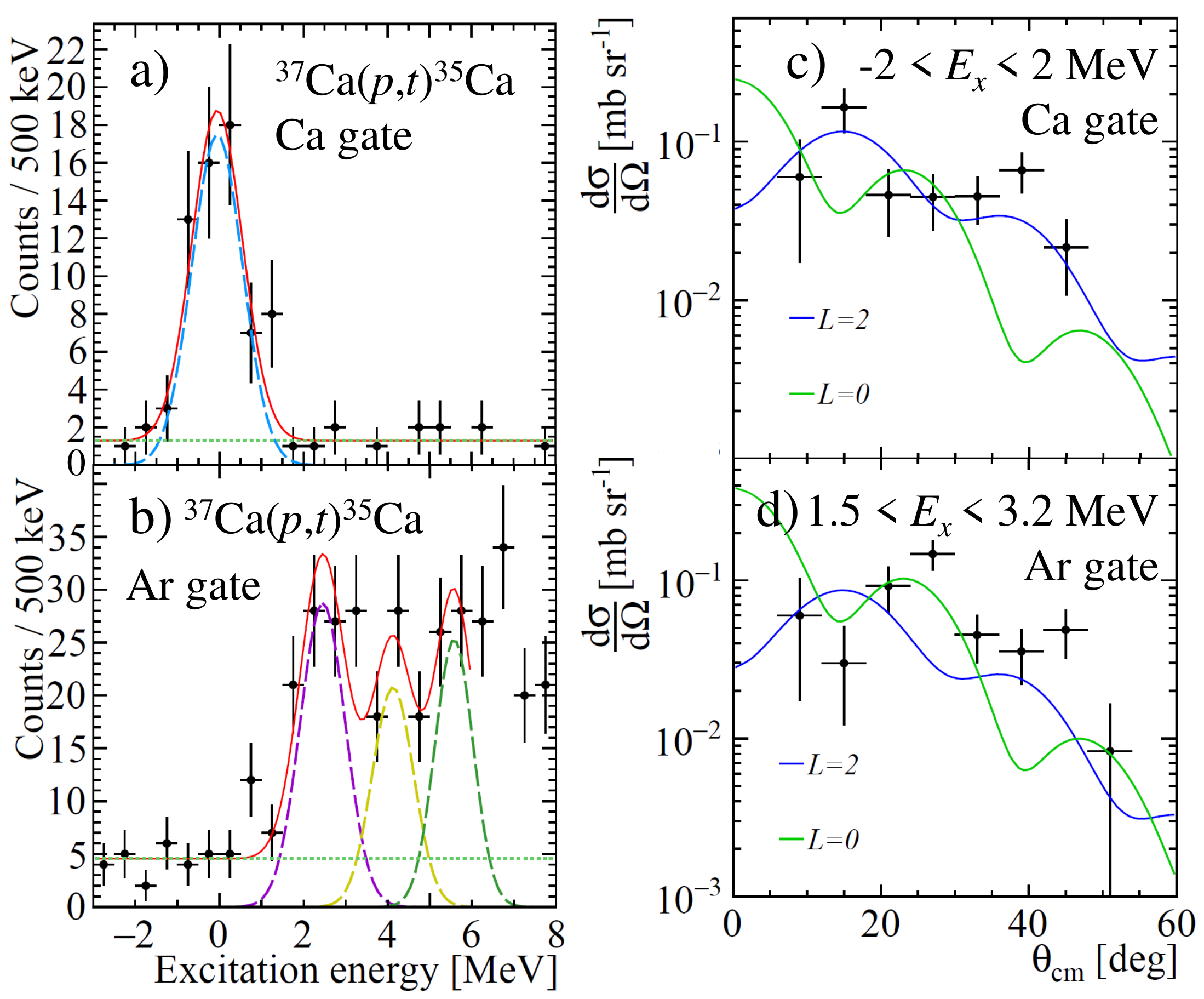}
\end{center}
\caption{a-b): Excitation energy spectrum of $^{35}$Ca obtained from the $^{37}$Ca$(p,t)$ transfer reaction with a gate on outgoing Ca a) and Ar b). The red lines show the best fit obtained while individual contributions are shown with different color. c-d):
	The experimental differential cross section obtained for the ground state c) and the first excited state d) identified in $^{35}$Ca using the $^{37}$Ca$(p,t)^{35}$Ca reaction is shown with the black points. A fit to the cross section is shown using the DWBA calculations performed for an $L=0$ (green) and an $L=2$ (blue) transfer.}
\label{fig:Res}
\end{figure}

\noindent {\sl Experimental techniques.}  The $^{37}$Ca nuclei were produced at about 50 MeV/nucleon by fragmentation reactions of a 95~MeV/nucleon $^{40}$Ca$^{20+}$ beam, with an average intensity  of about 2~e$\mu$A, in a 2-mm thick $^{9}$Be target. They were selected by the LISE3/GANIL spectrometer~\cite{Ann}, leading to a purity of 20\% and mean rate of 3$\times$10$^3$ pps.  They were identified by means of their time-of-flight (TOF) measurement between one of the two low-pressure {multiwire proportional chambers, CATS \cite{cats}, and the cyclotron radio frequency.  The two CATS detectors, placed at a relative distance of 51 cm, were also used to track the ions before their interaction with protons of the liquid hydrogen (LH) cryogenic target CRYPTA \cite{Koy20} of 9.7~mg$\,$cm$^{-2}$ at its center, placed at a distance of 67.8~cm downstream of the second CATS detector.  
	
	After their interaction with the protons of the LH  target, the outgoing ions were detected by a Zero Degree Detection (ZDD) system, composed of an ionization chamber (IC), used for their $Z$ identification, a set of two $XY$ drift chambers (DC), to determine their outgoing angles, and a thick plastic scintillator, mostly used for time-of-flight measurements. The energy and angle of the outgoing triton from the transfer reactions were determined by a set of six MUST2 telescopes \cite{must2}, covering angles between 3 and 37$^\circ$. Each of them is  composed of a 300-$\mu$m thick 10x10 cm double-sided silicon strip detector (DSSSD) with 128 strips on each side backed by sixteen 4-cm thick CsI crystals. The triton identification  was performed using their combined energy loss, $\Delta E$, measured in the DSSSD and  residual energy, $E$, measured in the CsI crystals. 
	
	The DSSSDs were calibrated strip by strip, using a triple-alpha source placed at the target position, covering an energy range from 5 to 6 MeV. The CsI detectors were calibrated using the kinematics of the tritons originating from $(p,t)$ reactions with incoming $^{38}$Ca and $^{36}$Ar nuclei, transmitted in another spectrometer setting, and for which $Q$-values are accurately known from precise mass measurements of the nuclei involved : $^{38}$Ca \cite{mass38Ca}, $^{36}$Ca \cite{mass36Ca},  $^{36}$Ar  \cite{mass36Ar} and $^{34}$Ar \cite{mass34Ar}. With this calibration, atomic masses and uncertainties can be determined from the weighted mean and standard deviation of four independent $Q$-value measurements performed in the four MUST2 telescopes, located at the closest distance from the target. A similar procedure was applied successfully  in Refs.~\cite{Lal21, PhD} to determine the atomic mass of $^{36}$Ca using the $^{37}$Ca($p,d$)$^{36}$Ca reaction. Moreover, a further confirmation of the method and of the calibration is obtained from the good agreement between the mass excess of $^{35}$K,  $\Delta M$ = -11205(110) keV, deduced in the present work using the known mass of $^{37}$K \cite{mass35K} and the precise measurement of $\Delta M$ = -11172.9(5) \cite{mass35K} in a penning trap.


	
	%
	%

\noindent {\sl Results.} The excitation energy $E_x$  of $^{35}$Ca, produced by the $^{37}$Ca($p,t$)$^{35}$Ca reaction, was deduced using the missing-mass method from the measurement of the energy and angle of the recoiling triton, detected in MUST2, and an incoming $^{37}$Ca identified and tracked in CATS. The $^{35}$Ca nucleus can be produced in a bound or unbound state. As the decay product $^{34}$K after one proton emission is unbound, the full excitation energy  spectrum of $^{35}$Ca can be obtained when gating on Ca Fig.~\ref{fig:Res} a)  or Ar Fig.~\ref{fig:Res} b) isotopes, identified through the measurement of their energy losses in the IC of the ZDD.  The red lines in Fig.~\ref{fig:Res} show the best fit obtained using multiple Gaussian functions plus a background contribution (green dashed line), generated by interactions of the beam particles with the windows of the LH$_2$ target and determined in a dedicated run with an empty target. The width of each peak used in the fit is constrained by simulations performed with the {\it nptool} package \cite{Matta2016}, the reliability of which is checked from the observed widths of isolated peaks in the reference reactions. The simulated width also matches the ground state value of 700~keV (sigma) in $^{35}$Ca, which is found to be the only bound state of $^{35}$Ca, as shown in Fig.~\ref{fig:Res}a).


The $Q$-value of the $^{37}$Ca($p,t$)$^{35}$Ca reaction has been determined from the energy of the ground state peak and the precise mass value of $^{37}$Ca \cite{mass38Ca}. This leads to a mass excess of $\Delta$M($^{35}$Ca) = 4777(105)~keV.}  Half of the uncertainty emerges from systematic effects such as the propagation of errors on the measured angle and energy of the tritons and on the energy calibration of the CsI detectors. The other half arises from the low statistics, about 15 counts per detector. 

The atomic mass of the ground state of $^{35}$Ca, which has isospin components $T_Z$ = -5/2, $T = 5/2$ and spin $J^{\pi} = 1/2^+$, can also be estimated from the Isobaric Multiplet Mass Equation (IMME) in its quadratic form with $T_Z$. 
A first attempt to derive the mass excess of  $^{35}$Ca from the IMME, leading to,  $\Delta$M($^{35}$Ca) = 4453(60)~keV, was obtained in \cite{Ays85}, which  deviates significantly from our value. However, by taking  accurate atomic masses of all nuclei involved (that were not available in 1985) and the energy of the $T$ = 5/2 isobaric analogue state in $^{35}$K proposed by Ref. \cite{Tri99}, we find $\Delta$M($^{35}$Ca$) = 4624(50)$~keV. This value agrees within one sigma with the present measurement. Conversely, the atomic mass of $^{35}$Ca extrapolated in the last Atomic Mass Evaluation compilation, 5190(200)~keV \cite{Wang21},  is 2$\sigma$ away from our result. 


The differential cross sections corresponding to the g.s. and to the first excited state of $^{35}$Ca are shown in Fig.~\ref{fig:Res}c,d). They have been obtained from the distribution of center-of-mass angles, after normalization of its amplitude using the number of incident nuclei, the density of protons in the target, as well as the detection efficiencies (intrinsic and geometrical) of the experimental setup. 
As the shape of the angular distribution is characteristic of the transferred angular momentum $L$,  DWBA calculations were performed with the code FRESCO \cite{fresco} assuming $L=0$ or $L=2$ angular momentum transfer, using the optical parameters given in \cite{Lal22Sup}. Both calculations of Fig.~\ref{fig:Res}c,d) consider only one dominant reaction channel,  that are, for $L=0$ (in green), the direct removal of a pair of neutrons from the $2s_{1/2}$ orbital and, for $L=2$ (in blue), the direct removal of one neutron from the $2s_{1/2}$ and the other from the $1d_{3/2}$ orbital. 

In spite of the limited statistics, it is seen that the shape of the ground state angular distribution of Fig.~\ref{fig:Res}c) is much better fitted when assuming an $L = 2$ transferred momentum ($\chi^2$/ndf = 10.2/7), rather than an $L = 0$ one ($\chi^2$/ndf = 27.1/7). This $L=2$ transfer from the 3/2$^+$ g.s. of $^{37}$Ca, corresponds to the removal of one neutron from the $2s_{1/2}$ and the other from the $1d_{3/2}$ orbital, leaving a single neutron in the $2s_{1/2}$ orbital. Therefore, the g.s. of $^{35}$Ca has a spin and parity of 1/2$^+$, which is in agreement with the established 1/2$^+$ g.s. spin value of the mirror nucleus $^{35}$P.

Excited states of $^{35}$Ca are visible in the excitation energy spectrum of Fig.~\ref{fig:Res}b), gated on outgoing Ar. There, the number of contributions used in the fit is guided by the statistical tests of the $\chi^2$ and the $p$-value, as well as the number of (3/2$^+$ and 5/2$^+$) states populated in the two-proton transfer quasi-mirror reaction $^{37}$Cl($^{11}$B,$^{13}$N)$^{35}$P \cite{Orr88}. The clear rising edge at about 2~MeV in Fig.~\ref{fig:Res}b) indicates the presence of the first excited state, which is found at 2.24(33)~MeV. However, due to the high density of states from 3~MeV onward and the present energy resolution, different fit functions lead to very similar $\chi^2/ndf$ (see Fig.~4.38 of Ref.~\cite{PhD}). This  precludes a conclusion about the number of higher excited states populated and their exact energies. It also significantly contributes to increase the uncertainty of the energy of the first excited state determined in this work. 

In the mirror nucleus $^{35}$P, the first excited state 3/2$^+$ at 2.3866(5)~MeV \cite{Wie08} has been strongly populated in the two-proton transfer reaction  $^{37}$Cl($^{11}$Be,$^{13}$N)$^{35}$P \cite{Orr88}, supporting its tentative  spin assignment in $^{35}$Ca. 
Given the large uncertainty on the centroid of the 3/2$^+$, one cannot bring valuable conclusions on the MED between the two nuclei.

The differential cross section of the 2.2~MeV excited state, shown in  Fig.~\ref{fig:Res}d),  was extracted using a condition on the excitation energy $1.5< E_x<3.2$~MeV and requiring an Ar isotope in the ZDD. The data are better fitted with an $L = 0$ transferred momentum ($\chi^2$/ndf = 20.4/7) than with $L = 2$ ($\chi^2$/ndf = 29.2/7). This favors spin and parity of $3/2^+$, with two holes in the neutron $2s_{1/2}$ orbital and one neutron in the $1d_{3/2}$ one.  Given the present energy resolution, one cannot exclude the contamination from a higher excited state, such as a $5/2^+$ state likely arising from the neutron $1d_{5/2}$ removal (as found at around 3.8 MeV in the mirror nucleus), that may account for the local maximum  at about 45$^\circ$.

\begin{figure}[!h]
\begin{center}
\centering
\includegraphics[width=0.8\columnwidth]{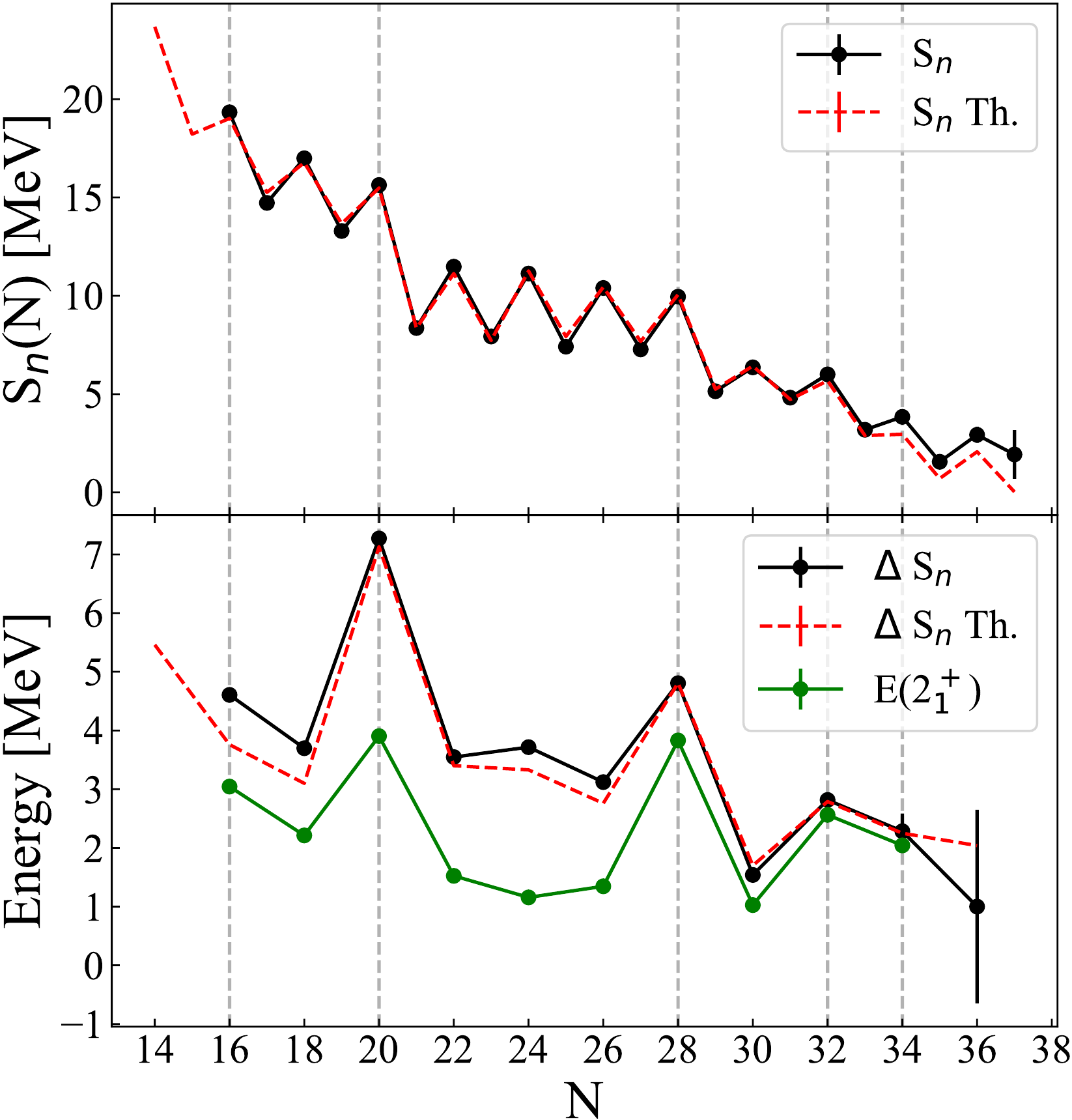}
\end{center}
\caption{Top: Experimental (black) and theoretical (red) one-neutron separation energy $S_n$ along the calcium isotopic chain. Bottom: $\Delta S_n$ and energy of the first $2^+$ excited state of even-even Ca isotopes. Theoretical $\Delta S_n$ values are shown in red.}
\label{fig:CaSyst}
\end{figure} 

\noindent {\sl Discussion.}- The one-neutron separation energy of $^{36}$Ca, $S_n(^{36}$Ca) = 19.331(110)~MeV, was determined by using the present mass measurement of $^{35}$Ca and the known value of $\Delta M(^{36}$Ca) = -6483.6 (56) keV \cite{mass36Ca}. The top panel of Fig.~\ref{fig:CaSyst} shows the experimental $S_n$ values along the calcium isotopic chain, including the new value of $^{36}$Ca. A pronounced decrease of $S_n$ is seen, on top of the odd-even oscillations, after having passed a  (sub-) shell gap, for $N$ = 16, 20 and 28, as well as  $N$=32 and 34 \cite{Wien13, Mich18}. 

The bottom part of Fig.~\ref{fig:CaSyst} displays differences of one-neutron separation energies, $\Delta S_n(N) = S_n(N) - S_n(N+1)$ for even-even Ca isotopes.  At closed shells, where the effect of pairing is significantly reduced, $\Delta S_n(N)$ is directly related to the amplitude of the shell gap. The maximum of magicity along the Ca isotopic chain is reached at $N=20$ ($\sim 7.20$~MeV) for the self-conjugate $^{40}$Ca nucleus. At $N=16$, a shell gap of 4.61(11)~MeV is obtained, the size of which is very similar to the one at $N = 28$ ($\sim 4.80$~MeV), significantly larger than that at $N=32$ ($\sim 2.82$~MeV) and twice at large as at $N=$34 ($\sim 2.28$~MeV).  This provides a strong evidence of the magicity at $N = 16$, corroborated by the systematics of  first $2^+$ excited states, also presented in the bottom part of Fig.~\ref{fig:CaSyst}, which follows the same trend as the $\Delta S_n(N)$ values. Note that the spacing between these two curves ($\Delta S_n(N)$ and $2^+$) is weaker at $N = 32$ and $N = 34$ than for other magic shells. This is likely due to the fact that the $2^+$ states at low energy are more of a pure neutron origin and coincide with the amplitude of the neutron gap, while those at higher energy combine neutron and proton excitations and are more subject to correlations.  


Shell-model (SM) calculations have been carried out with the Antoine \cite{rmp} code using the {\sl sdpf} valence space below $A=41$ and the {\sl pf} one from $A=41$ onward.
The nuclear, isospin conserving parts, are given by the {\sl sdpf-u-mix} interaction \cite{caurier2014} and by the {\sl pfsdg-u} interaction \cite{pfsdg}, respectively.
The two-body matrix elements of the Coulomb interaction are computed with harmonic oscillator wave functions with \mbox{$\hbar \omega =  41 A^{-1/3} - 25 A^{-2/3}$}. The Coulomb corrections to the single-particle energies are taken from the experimental spectra of the $A$ = 17 and $A$ = 41 mirror nuclei. 

The theoretical $S_n(N)$  and $\Delta S_n(N)$ values are shown in red in Fig.~\ref{fig:CaSyst}. The overall $S_n$ trend is found to be well reproduced by SM calculations. In particular, the theoretical $\Delta S_n(N)$ values are in good agreement with the experimental ones at $N = 20,28,32$ and 34. The predicted amplitude of the $N = 16$ shell gap, $\Delta S_n = 3.77$~MeV, is however 840(110) keV lower than the  experimental value of 4.61(11)~MeV.  Shell gaps of 3.84 and 4.00 MeV, predicted by the USDA and USDB interactions \cite{USD}, are closer but still smaller than the experimental value. In fact this discrepancy can be attributed to a residual defect of the USD family of interactions (notice that USD is the {\it sd} part of {\sl sdpf-u-mix})  that produce a somewhat (20\%) smaller T=1  $(1s_{1/2})^2$ monopole interaction. The $\Delta S_n$ value at $N=16$ measured in this work is compatible to the one of $\Delta S_p$ at $Z=16$ in $^{36}$S of $4.709$~MeV, indicating that the mirror symmetry conserves the size of the 16 gap. Our SM calculations underestimate the $Z=16$ gap in $^{36}$S by the same amount.

The theoretical values of the excitation energy of the $3/2^+$ state in $^{35}$Ca, $E_x$=2.38~MeV obtained with the {\sl sdpf-u-mix} interaction, is in good agreement with the experimental ones of $E_x=2.24(33)$~MeV, further supporting its spin-parity assignment. The associated MED between $^{35}$Ca and $^{35}$P is predicted to be of -300~keV, compatible with the shifts of about $-250$~keV of the 1$^+$ and 2$^+_1$ states of the $^{36}$Ca - $^{36}$S mirror pair~\cite{Lal22}, suggesting a similar origin. Shell model calculation predicts the next shell closure in the Ca isotopes to be at $N = 14$, with a large gap of 5.46~MeV in  unbound $^{34}$Ca.

\begin{figure}[!h]
\begin{center}
\centering

\includegraphics[width=0.8\columnwidth]{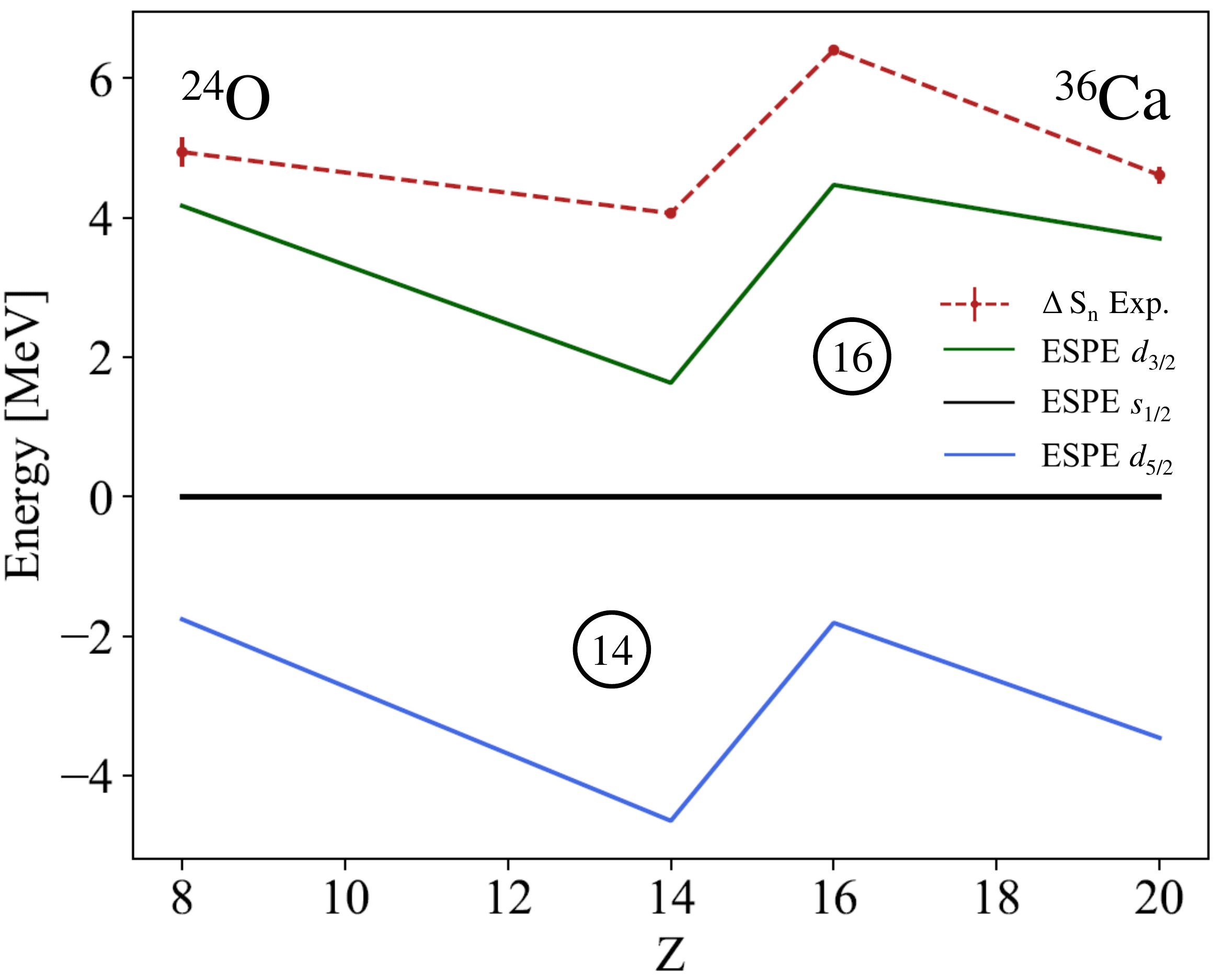}    

\end{center}
\caption{The effective single particle energies (ESPE)  of the neutron $d_{5/2}$, $s_{1/2}$ and $d_{3/2}$ orbitals at $N=16$ in the {\sl sdpf-u-mix} interaction are shown with the full colored line, relative to the $s_{1/2}$ one. The red dashed lines shows experimental $\Delta S_n$ values along $N=16$ at sub-shell closures. }
\label{fig:ESPE}
\end{figure}

The $N = 16$ gap is almost equally large at both edges of the nuclear chart: it amounts to  $\Delta S_n =4.61(11)$~MeV in $^{36}$Ca, which lies close to the proton drip line, and to $\Delta S_n =4.94(20)$~MeV in $^{24}$O, which is the last bound oxygen isotope.   By applying the $A^{-1/3}$ factor related to the overall compression of the level spacing in atomic nuclei with increasing $A$, one finds that the $N=16$ gap is larger in $^{36}$Ca than in $^{24}$O, viz. 4.61 $\times$ (36/24)$^{1/3}$= 5.26(12) MeV in $^{36}$Ca as compared to 4.94(20) in $^{24}$O.  

Fig~\ref{fig:ESPE} shows the evolution of the effective single particles energies (ESPE)  of the neutrons $1d_{5/2}$ (blue), $2s_{1/2}$ (black) and $1d_{3/2}$ (green) orbitals at $N=16$, computed with the {\sl sdpf-u-mix} interaction, as a function of the proton number. The ESPE of the $1d_{5/2}$ and $1d_{3/2}$ orbitals, given relative to the $2s_{1/2}$ one, directly allows to determine the predicted amplitudes of the $N=14$ and $N=16$ gaps, respectively. Starting from  $^{24}$O, with the addition of 6 protons to the $1d_{5/2}$ orbit, the attractive tensor interaction between the two spin-orbit partners reduces the predicted single-particle $N=16$ gap by about 2.5 MeV at $Z=14$. The addition of only two more protons to the $2s_{1/2}$ orbit restores the gap at $Z=16$, while adding 4 more to the $1d_{3/2}$ proton orbital, brings it back close to the value in $^{24}$O. 

This predicted evolution of the $N=16$ gap (green line) is compared in Fig~\ref{fig:ESPE} to experimental $\Delta S_n$ values (red dashed line). The  $\Delta S_n$ values, that corresponds to experimental correlated gaps, are systematically larger than the size of $N=16$ derived from ESPE. However, the overall trend is similar. It is worth noting that, in spite of a very large gap, $^{32}$S does not show significant signs of magicity because of the enhancement of the pairing and quadrupole correlations for this $N = Z$ nucleus.

\noindent {\sl Conclusion.}
The $^{37}$Ca($p,t$)$^{35}$Ca reaction was performed to obtain the first measurement of the atomic mass of $^{35}$Ca, $\Delta M(^{35}$Ca) = 4.777(105)~keV, as well as the excitation energy of its first excited state at 2.24(33)~MeV. The measured differential cross sections together with our shell model calculations support a spin parity of $1/2^+$ for the ground state and $3/2^+$ for the first excited state. The atomic mass was used to infer the amplitude of the gap at $N = 16$ in $^{36}$Ca, 4.61(11)~MeV, which is very similar to the one of $N = 28$, significantly larger than that at $N=32$ ($\sim 2.82$~MeV) and twice at large as at $N=$34 ($\sim 2.28$~MeV). This result corroborates the arguments in favor of double magicity of $^{36}$Ca from the observation of high-energy $2^+$ and $1^+$ states and  their large neutron-removal spectroscopic factors $C^2S$ obtained from $^{37}$Ca($p,d$)$^{36}$Ca~\cite{Lal22}. The magicity of $^{36}$Ca  is further confirmed by  its small charge radius~\cite{Mill19}, the smallest among all Ca isotopes.

The amplitude of the $N = 16$ gap in $^{36}$Ca is also comparable to the one found in $^{24}$O, described as a doubly-magic nucleus (see, e.g. \cite{Hoff09, Tsho12}). The fact that $N = 16$ magicity is strongly present at both edges of the nuclear binding gives strong constraints to the proton-neutron interactions involved when adding 12 protons to the $sd$ shells. Thus far, the use of state-of-the-art shell-model interactions underestimate  the gap derived from $\Delta S_n$ by about 0.61 (USDB) to 0.84 MeV ({\sl sdpf-u-mix}). The SM calculation carried out with both interactions predicts the next shell closure of the Ca isotopic chain to be at $N=14$ in the doubly-magic and unbound $^{34}$Ca, which is the mirror of the bubble nucleus $^{34}$Si \cite{Muts17}.

\acknowledgments {\small
The continued support of the staff of the GANIL facility is gratefully acknowledged. DS was supported by the JSPS KAKENHI Grant Number 19H01914. 
AP's work is supported in part by  grants CEX2020-001007-S  funded by MCIN/AEI/10.13039/501100011033 and 
PID2021-127890NB-I00. Support from the NFS grant PHY-2110365 is also acknowledged.}

\end{document}